\documentclass[prl,twocolumn,showpacs,preprintnumbers,amsmath,amssymb]{revtex4}
\usepackage{graphicx}
\begin{document}

\noindent {\bf  Comment on ``Non-Mean-Field Behavior of the Contact Process on
Scale-Free Networks''}\\

Recently, Castellano and Pastor-Satorras~\cite{CPS06} used the finite size
scaling (FSS) theory to analyze simulation data for the contact process (CP) on
scale-free networks (SFNs) and claimed that its absorbing critical behavior is
not consistent with the mean-field (MF) prediction. Furthermore, they pointed
out large density fluctuations at highly connected vertices  as a possible
origin for non-MF critical behavior. In this Comment, we propose a scaling
theory for relative density fluctuations in the spirit of the MF theory, which
turns out to explain simulation data perfectly well. We also measure the value
of the critical density decay exponent $\beta/\nu_\|$, which agrees well with
the MF prediction. Our results strongly support that the CP on SFNs still
exhibits a MF-type critical behavior.

\begin{figure}[b]
\includegraphics[width=0.9\columnwidth]{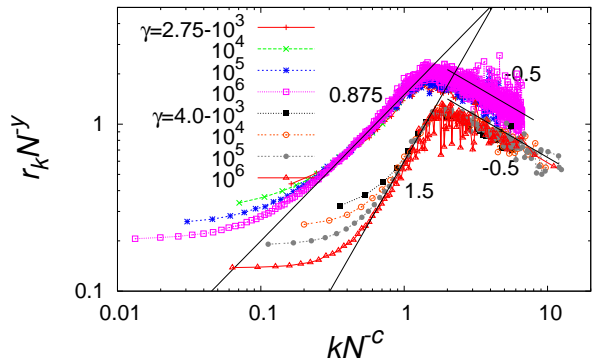}
\caption{(Color online) Data collapse of the relative density fluctuations
$r_k$ averaged over surviving samples in the steady state for
$\gamma=2.75~(p_c=0.4240,~c=0.363,~y=0.104)$ and
$\gamma=4.0~(p_c=0.3570,~c=0.25,~y=0.125)$, respectively.} \label{fig1}
\end{figure}

In contrast to equilibrium models with thermal noise, nonequilibrium models
exhibiting absorbing critical phenomena involve so-called multiplicative noise.
In particular, the noise amplitude is proportional to the square root of the
particle density in models belonging to the directed percolation class such as
CP~\cite{DP}. Therefore, at the MF level neglecting spatial fluctuations, the
particle number fluctuation should be proportional to the square root of the
total number of particles. It is easy to derive the relative density
fluctuation on vertices of degree $k$ as
\begin{equation}
r_k = \Delta \rho_k/\rho_k\sim [N \rho_k P_k ]^{-1/2}, \label{eq1}
\end{equation}
where $\rho_k$ is the partial density on vertices of degree $k$, $N$ is the
total number of vertices, and $P_k$ is the degree distribution. As $\rho_k\sim
\rho k$ near criticality ($\rho$: total density), we find, for reasonably large
$k$,
\begin{equation}
r_k\sim (N \rho)^{-1/2}k^{(\gamma-1)/2}, \label{eq2}
\end{equation}
where $\gamma$ is the degree exponent.

For large enough $k>N^{1/\gamma}$ where the degree distribution becomes almost
flat in practice ($NP_k\sim O(1)$), we expect $r_k\sim (\rho k)^{-1/2}$. With
these two limiting results and also with $\rho\sim N^{-\alpha}$ at criticality,
one may assume a critical scaling relation as
\begin{equation}
r_k= N^{(\alpha-1/\gamma)/2} f(kN^{-1/\gamma}), \label{eq4}
\end{equation}
where $f(x)\sim x^{(\gamma -1)/2}$ as $x\rightarrow 0$ and $f(x)\sim x^{-1/2}$
as $x\rightarrow\infty$. In Fig.~\ref{fig1}, our simulation results at
criticality for $\gamma=2.75$ and 4.0 with various sizes $N=10^3, \cdots, 10^6$
show excellent agreements with our prediction. Note that the case of
$\gamma=4.0$ where the $\gamma$-independent ordinary MF appears also shows the
similar behavior in $r_k$.

\begin{figure}[t]
\includegraphics[width=0.9\columnwidth]{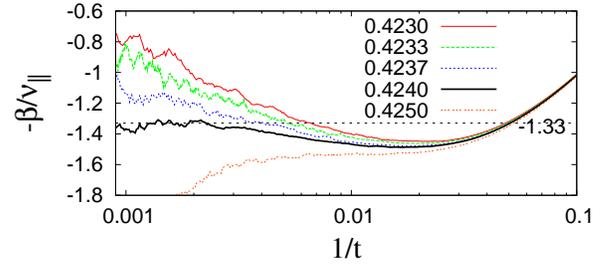}
\caption {(Color online) Effective exponents $\beta/\nu_{||}$ for $\gamma=2.75$
with $N=10^7$. All samples are alive when data are obtained.} \label{fig2}
\end{figure}

We also observe the density decay dynamics for $\gamma=2.75$ in
Fig.~\ref{fig2}, where the effective decay exponents $\beta/\nu_\|$ are plotted
against $1/t$. Our best estimate is $\beta/\nu_\|=1.34(5)$ which agrees well
with the MF prediction of $\nu_\|=1$ and $\beta=1/(\gamma-2)$ for $\gamma<3$.

Finally, we comment on the FSS exponent $\alpha=\beta/{\bar\nu}$. Vertices of
degree $k>N^{1/\gamma}$ are practically irrelevant. Accordingly, the cutoff in
$k$ bigger than $N^{1/\gamma}$ can not influence critical FSS which should be
determined by the competition between correlated volume and system size. We
conjecture that $\bar\nu=(\gamma-1)/(\gamma-2)$ from the hyperscaling-type
argument~\cite{HHP}. Our numerical estimate $\beta/{\bar\nu}=0.59(2)$ for
$\gamma=2.75$ agrees well with our conjecture. In fact, the results of
Ref.~\cite{CPS06} also agree reasonably well with our conjecture. In
conclusion, we believe that the CP on SFNs exhibits the MF-type critical
behavior.

\acknowledgments This work was supported in part by the KRF Grant (MOEHRD)
(R14-2002-059-01000-0). \vspace{0.3cm}

\noindent
Meesoon Ha$^{1,2}$, Hyunsuk Hong$^2$, and Hyunggyu Park$^{1}$ \\
{\small
$^1$School of Physics, Korea Institute for Advanced Study\\
\hphantom{$^1$}Seoul 130-722, Korea\\
$^2$Department of Physics and RIPC \\
\hphantom{$^2$}Chonbuk National University, Jeonju 561-756, Korea\\}

\noindent\today

\itemize
\bibitem{CPS06} C. Castellano and R. Pastor-Satorras, Phys. Rev. Lett.
{\bf 96}, 038701 (2006).
\vspace{-2.7mm}
\bibitem{DP} H. Hinrichsen, Adv. Phys. {\bf 49}, 815 (2000).
\vspace{-2.7mm}
\bibitem{HHP} H. Hong, M. Ha, and H. Park (in preparation).

\end{document}